\documentclass[11pt]{amsart}

\usepackage{amsmath,amssymb,latexsym,amsthm,enumerate}
\usepackage{verbatim}
\usepackage{hyperref}

\theoremstyle{plain}
\newtheorem{theorem}{Theorem}%[section]

\newtheorem{lemma}[theorem]{Lemma}

\theoremstyle{definition}

\theoremstyle{remark}

\newtheorem*{remark}{Remark}
\newtheorem*{remarks}{Remarks}

\newcounter{numpar}[section]
\newcommand*{\wt}{\widetilde}

\newcommand*{\dd}{\mathrm d}     %For dx, dy, etc.

\newcommand*{\cD}{\mathcal D}

\newcommand*{\cF}{\mathcal F}

\newcommand*{\bbR}{\mathbb R}

\newcommand*{\EE}{\mathsf E}
\newcommand*{\PP}{\mathsf P}

\newcommand*{\E}{\mathrm{e}}

\begin{document}
\title{A note on essential smoothness in the Heston model}

\author{Martin Forde}
\address{Department of Mathematical Sciences, Dublin City University, Ireland} 
\email{martin.forde@dcu.ie}

\author{Antoine Jacquier}
\address{Department of Mathematics, TU Berlin, Germany} 
\email{jacquier@math.tu-berlin.de}

\author{Aleksandar Mijatovi\'{c}}
\address{Department of Statistics, University of Warwick, UK}
\email{a.mijatovic@warwick.ac.uk}



\begin{abstract}
This note studies an issue relating to essential smoothness that can arise when the theory of large
deviations is applied to a certain option pricing formula
in the Heston model.
%This note studies a subtle issue 
%with essential smoothness
%that can arise when 
%the theory of large deviations 
%is applied to
%a certain 
%option pricing formula. 
The note identifies a gap,
based on this issue,
in the proof of Corollary~2.4 in~\cite{FJ09Large} and 
describes how to circumvent it.
%this gap can be avoided. 
This completes the proof of 
Corollary~2.4 in~\cite{FJ09Large} and hence of the
main result in~\cite{FJ09Large}, 
which describes the limiting behaviour
of the implied volatility smile in the Heston model far from maturity. 
%Furthermore the note provides a foundation for the large deviation theory 
%approach to the study of the implied volatility asymptotics at large maturity
%via the option pricing formula. 
\end{abstract}

\maketitle

%========================================================================================
\section{Introduction}
\label{sec:intro} 

In~\cite{FJ09Large} the authors study the limiting behaviour of the implied volatility  
in the Heston model as maturity tends to infinity. 
%The main result in~\cite{FJ09Large},
%which gives a formula for the limit of the implied volatility at an appropriately scaled 
%time-dependent strike, is of significant interest for two reasons: (a) the limit itself
%is an important quantity that can be expressed explicitly in terms of the model
%parameters and can therefore facilitate the calibration of the model and (b) the
%derivation of the limit proposed in~\cite{FJ09Large} is probabilistic in nature
%as it provides an interesting application of the theory of large deviations
%via the option pricing formulae in~\eqref{eq:Put_Formulae} and~\eqref{eq:Call_Formulae}.
The main aim of this note is to give a rigorous account 
of the relationship between the concept of essential smoothness and the large deviation 
principle for the family of random variables
$(X_t/t\pm E_\lambda/t)_{t\geq1}$,
where the process 
$X$
denotes the log-spot in Heston model~\eqref{eq:Heston_SDE}
and 
$E_\lambda$
is an exponential random variable with 
parameter
$\lambda>0$
independent of 
$X$.
%(see Lemma~\ref{lem:perturbation} and Theorem~\ref{thm:Fix}).
%and the option pricing formulae
%of~\eqref{eq:Put_Formulae} and~\eqref{eq:Call_Formulae}
%in the context of~(b). 
%The motivation for this work is twofold: one, the connection between large 
%deviations and pricing theory 
%via~\eqref{eq:Put_Formulae} and~\eqref{eq:Call_Formulae} 
%is of independent interest and Lemma~\ref{lem:perturbation}
%and Theorem~\ref{thm:Fix} pave the way for similar applications in models
%other than the Heston stochastic volatility model; and two, 
This note fills a gap in the proof of Corollary~2.4 in~\cite{FJ09Large} 
and hence completes the proof of the main result in~\cite{FJ09Large},
which describes the limiting behaviour of the implied volatility smile in the 
Heston model far from maturity. 

The note is organized as follows. %In Section~\ref{sec:formula}
%pricing formulae~\eqref{eq:Put_Formulae} and~\eqref{eq:Call_Formulae}
%are stated and proved. 
Section~\ref{sec:trap} describes the relevant concepts
of the large deviation theory and discusses how  
the effective domain changes when a family of random variables is perturbed
by an independent exponential random variable. Section~\ref{sec:Trap}
discusses the failure of essential smoothness when the Heston model 
is perturbed by an independent exponential, which is what causes the 
gap in the proof of Corollary~2.4 in~\cite{FJ09Large}. 
Section~\ref{sec:Trap}
also proves Theorem~\ref{thm:Fix}, which fills the gap. 
\section{The large deviation principle for random variables in $\bbR$}
\label{sec:trap}

We briefly recall the basic facts of the large deviation theory 
in
$\bbR$
(see monograph~\cite[Ch. 2]{DemboZeitouni} for more details).
Let  
$(Z_t)_{t\geq1}$
be a family of random variables
%taking values in
with
$Z_t\in\bbR$.
%Recall that 
%$J:\bbR\to(-\infty,\infty]$
%is \textit{lower semicontinuous} 
%if 
%$\{x:J\left(x\right)\leq\alpha\}$
%is closed in 
%$\bbR$ for any
%$\alpha\in\bbR$
%(intuitively 
%for any 
%$x_0\in\bbR$
%the values of 
%$J$
%near 
%$x_0$
%are either close to 
%$J(x_0)$
%or greater than 
%$J(x_0)$).
$J$
is a \textit{rate function}
if it is lower semicontinuous
and
%nonnegative (i.e. 
$J(\bbR)\subset[0,\infty]$
holds.
%and
%$J:\bbR\to[0,\infty]$
%a \textit{lower semicontinuous}
%function
%(i.e. %the set
%$\{x:J\left(x\right)\leq\alpha\}$
%is closed in 
%$\bbR$ for any
%$\alpha\in[0,\infty)$).
The family
$(Z_t)_{t\geq1}$
satisfies the \textit{large deviation principle (LDP)}
with the \textit{rate function}
$J$
if for every Borel set
$B\subset\bbR$
we have
\begin{equation}
\label{eq:DefLDP}
-\inf_{x\in B^\circ}J(x)\leq\liminf_{t\to\infty}\frac{1}{t}\log \PP\left[Z_t\in B\right]
\leq
\limsup_{t\to\infty}\frac{1}{t}\log \PP\left[Z_t\in B\right]\leq-\inf_{x\in \overline B}J(x),
\end{equation}
with the convention
$\inf\emptyset =\infty$
%Throughout this note
the relative notions of 
interior
(interior $B^\circ$,
closure
$\overline B$
and boundary
$\overline B\setminus B^\circ$
are
%are taken with respect to
in the topology of
$\bbR$).

%where 
%$\inf\emptyset =\infty$
%and
%the interior
%$B^\circ$
%and closure
%$\overline B$
%%of 
%%$B$
%are taken in 
%the topology of
%$\bbR$.

%Does
%a given 
%a family 
%$(Z_t)_{t\geq1}$
%satisfy the LDP and if so, what is its rate function?
The G\"artner-Ellis theorem 
(Theorem~\ref{thm:GartnerEllis} below)
gives 
%an affirmative answer 
%to the question under general 
sufficient conditions 
for 
a family 
$(Z_t)_{t\geq1}$
to satisfy the LDP
(see monograph~\cite[Section 2.3]{DemboZeitouni}
for details).
Let 
$\Lambda_t(u):=\log \EE\left[\E^{uZ_t}\right]\in(-\infty,\infty]$
be the cumulant generating function
of
$Z_t$.
Assume that for every
$u\in\bbR$
%the following limit
%exists 
%as an extended real number
%in
%$(-\infty,\infty]$
\begin{eqnarray}
\label{eq:LDP_Assumption}
\Lambda(u)  :=  \lim_{t\to \infty}\Lambda_t(tu)/t
\quad\text{exists in } %as an extended real number}
%\quad
[-\infty,\infty]
\qquad\text{and}\qquad
 0  \in  \cD_\Lambda^\circ, 
%&\text{where}
%\cD_\Lambda:=\{u\in\mathbb R\>:\>|\Lambda(u)|<\infty\}.
\end{eqnarray}
where
$\cD_\Lambda:=\{u\in\mathbb R:\Lambda(u)<\infty\}$
is the \textit{effective domain} of
$\Lambda$
and
$\cD_\Lambda^\circ$
is its interior.
%H\"older's inequality implies that 
%$u\mapsto\Lambda_t(tu)/t$
%is convex for every
%$t\geq1$
%and hence  so is
%$\Lambda$
%by~\cite[Theorem~10.8]{Rockafellar}.
%Since 
%$\Lambda(0)=0$,
%the convexity of
%$\Lambda$
%and
%$ 0\in\cD_\Lambda^\circ$
%imply
%$\Lambda(u)>-\infty$
%for all
%$u\in\bbR$.
The \textit{Fenchel-Legendre transform}
$\Lambda^*$
of the convex function 
$\Lambda$
is
defined by the formula
\begin{eqnarray}
\label{eq:DefFenchelLegendreTransf}
\Lambda^*(x) &:=& \sup\{ux-\Lambda(u)\>:\>u\in\bbR\}
\quad\text{for}\quad
x\in\bbR.
\end{eqnarray}
%with an effective domain
%$\mathcal D_{\Lambda^*}:=\{x\in\mathbb R\>:\>\Lambda^*(x)<\infty\}$.
Under the assumption in~\eqref{eq:LDP_Assumption},
$\Lambda^*$
is
lower semicontinuous with compact level
sets
$\{x:\Lambda^*(x)\leq \alpha\}$
(see~\cite[Lemma 2.3.9(a)]{DemboZeitouni})
and
$\Lambda^*(\bbR)\subset[0,\infty]$
and hence satisfies the definition of 
a \textit{good rate function}.
%
%the definition in~\eqref{eq:DefFenchelLegendreTransf} 
%implies that
%$\Lambda^*$
%is
%lower semicontinuous
%(since it is a supremum of continuous functions),
%$\Lambda^*(\bbR)\subset[0,\infty]$
%(since
%$\Lambda(0)=0$)
%and the level sets
%$\{x:\Lambda^*(x)\leq \alpha\}$
%are compact for all
%$\alpha\geq0$
%(see~\cite[Lemma 2.3.9(a)]{DemboZeitouni}),
%i.e. 
%$\Lambda^*$
%satisfies the definition of 
%a \textit{good rate function}.
%It also follows from~\eqref{eq:DefFenchelLegendreTransf} 
%that 
%$\Lambda^*$
%is convex since it is a supremum of linear functions.
We now state the G\"artner-Ellis theorem
(see~\cite[Section 2.3]{DemboZeitouni} for its proof).

%\begin[G\"artner-Ellis theorem]{theorem}
\begin{theorem}
\label{thm:GartnerEllis}
Let the random variables
$(Z_t)_{t\geq1}$
%be a family of random variables
%that 
satisfy the assumption in~\eqref{eq:LDP_Assumption}.
If
$\Lambda$
is essentially smooth and lower semicontinuous, then LDP
holds for
$(Z_t)_{t\geq1}$
with the good rate function
$\Lambda^*$.
\end{theorem}

The function
$\Lambda:\bbR\to(-\infty,\infty]$
defined in~\eqref{eq:LDP_Assumption}
is \textit{essentially smooth}
if it is (a)~differentiable in
$\mathcal D^\circ_\Lambda$
and (b)~\textit{steep},
i.e.
$\lim_{n\to\infty}|\Lambda'(u_n)|=\infty$
for every sequence
$(u_n)_{n\in\mathbb N}$
in
$\mathcal D^\circ_\Lambda$
that converges to a boundary point of
$\mathcal D^\circ_\Lambda$. 
%Since the notions of boundary and interior 
%are with respect to the ambient topology of 
%$\bbR$, 
%the assumption of essential smoothness in 
%Theorem~\ref{thm:GartnerEllis}
%is automatically satisfied if
%$\cD_\Lambda=\bbR$. 
If 
$\cD_\Lambda^\circ$ 
is a strict subset of 
$\bbR$,
which is the case in the setting of~\cite{FJ09Large}
(see also Section~\ref{sec:Trap} below),
essential smoothness, which
plays a key role in the proof of 
Theorem~\ref{thm:GartnerEllis},
is not automatic. 
%In this case it 
%plays an important role in %the proof of 
%Theorem~\ref{thm:GartnerEllis}
%as it facilitates 
%a key step in its proof based on an intricate convex analysis
%argument given in~\cite[Lemma~2.3.12]{DemboZeitouni}.

%The asymptotic behaviour of option prices in~\cite{FJ09Large}
%is based on the formulae in~\eqref{eq:Put_Formulae}
%and~\eqref{eq:Call_Formulae}.
The following question
is of central importance in~\cite{FJ09Large}: does
the LDP persist if a family of random variables
$(Z_t)_{t\geq1}$
is perturbed by an independent exponential random variable
$E_1$?
It is implicitly assumed in the proof of Corollary~2.4 in~\cite{FJ09Large} 
(see the last line on page~17 and lines~4 and~14 on page~18)
that if 
$(Z_t)_{t\geq1}$
satisfies the assumptions of Theorem~\ref{thm:GartnerEllis},
then so do the families 
$(Y^{1+}_t)_{t\geq1}$
and
$(Y^{1-}_t)_{t\geq1}$,
where
$Y_t^{1\pm}=Z_t\pm E_1/t$,
%$(Z_t+E_1/t)_{t\geq1}$
%and
%$(Z_t-E_1/t)_{t\geq1}$
and the LDP is applied.
In particular 
the authors in~\cite{FJ09Large}
assume that the limiting cumulant generating
functions of 
$(Y^{1\pm}_t)_{t\geq1}$
are essentially smooth. However the following simple lemma holds.

\begin{lemma} 
\label{lem:perturbation} 
Let $(Z_t)_{t\geq1}$ satisfy the assumption in~\eqref{eq:LDP_Assumption}
with a limiting cumulant generating function
$\Lambda$.
%be a family of random variables and let $\Lambda$
%be the corresponding cumulant generating function 
%defined in~\eqref{eq:LDP_Assumption}
%with an effective domain
%$\mathcal D_\Lambda$
%and a Fenchel-Legendre transform 
%$\Lambda^*$.
%Assume further that 
%$\Lambda$
%is differentiable in
%$\mathcal D_\Lambda^\circ$.
Let
$\lambda>0$
and
$E_\lambda$
an exponential random variable
independent of 
$(Z_t)_{t\geq1}$
with
$\EE[E_\lambda]=1/\lambda$
and let
$Y_t^{\lambda\pm}:=Z_t\pm E_\lambda/t$.
Then the families of random variables
$(Y^{\lambda\pm}_t)_{t\geq1}$
satisfy the assumption in~\eqref{eq:LDP_Assumption}
and the corresponding limiting cumulant generating functions
%~\eqref{eq:LDP_Assumption} 
%and their Fenchel-Legendre transforms~\eqref{eq:DefFenchelLegendreTransf} 
are given by
\begin{align*}
\Lambda^{\lambda+}(u) & = 
\left\{
\begin{array}{ll}
\displaystyle
\Lambda(u),
& \text{if } u\in \mathcal D_\Lambda\cap(-\infty,\lambda),\\
\infty,
& \text{otherwise}, 
\end{array}
\right.
\quad 
\text{and}\quad
%\Lambda_{Y^+}^*(x)
%= 
%\left\{
%\begin{array}{ll}
%\displaystyle
%\Lambda^*(x),
%\quad & \text{if } x<\Lambda_-'(1),\\
%x-\Lambda_-(1),
%\quad & \text{if } x\geq\Lambda_-'(1),
%\end{array}
%\right.
%\\
\Lambda^{\lambda-}(u) & = 
\left\{
\begin{array}{ll}
\displaystyle
\Lambda(u),
& \text{if } u\in \mathcal D_\Lambda\cap(-\lambda,\infty),\\
\infty,
& \text{otherwise}. 
\end{array}
\right.
%\quad \text{and}\quad
%\Lambda_{Y^-}^*(x)
%= 
%\left\{
%\begin{array}{ll}
%\displaystyle
%\Lambda^*(x),
%\quad & \text{if } x>\Lambda_+'(-1),\\
%-x-\Lambda_+(-1),
%\quad & \text{if } x\leq\Lambda_+'(-1).
%\end{array}
%\right.
\end{align*}
%If 
%the effective domain
%$\mathcal D_\Lambda$
%is contained in
%$(-\infty,1)$
%(resp.
%$(-1,\infty)$)
%the values
%$\Lambda'_-(1)$
%and
%$\Lambda_-(1)$
%(resp.
%$\Lambda'_+(-1)$
%and
%$\Lambda_+(-1)$)
%are taken to be
%$\infty$
%(resp.
%$-\infty$).
%Otherwise they are defined as
%$\Lambda'_-(1):=\lim_{u\nearrow1}\Lambda'(u)$,
%$\Lambda_-(1):=\lim_{u\nearrow1}\Lambda(u)$
%and
%$\Lambda'_+(-1):=\lim_{u\searrow-1}\Lambda'(u)$,
%$\Lambda_+(-1):=\lim_{u\searrow-1}\Lambda(u)$
%where the limits exist as extended real numbers.
\end{lemma}

\begin{remarks}\noindent \textbf{(a)} Let 
$(Z_t)_{t\geq1}$
satisfy the assumption in~\eqref{eq:LDP_Assumption}
and assume further that 
$\Lambda$
is differentiable in
$\cD_\Lambda^\circ$.
If 
$1\in\cD_\Lambda^\circ$,
then the right-hand boundary point of
the interior of the effective domain
$\cD_{\Lambda^{1+}}^\circ$
is equal to 
$1$
and 
Lemma~\ref{lem:perturbation}
implies that
the limiting cumulant generating function
$\Lambda^{1+}$
of 
$(Y^{1+}_t)_{t\geq1}$
is
\begin{itemize}
\item 
neither essentially smooth,
since
$\Lambda^{1+}$
is not steep at
$1$, 
\item 
nor
lower semicontinuous at 
$1$,
since it is differentiable in
$\cD_{\Lambda^{1+}}^\circ$
with
%and takes value 
$\Lambda^{1+}(1)=\infty$.
\end{itemize}
Loss of steepness and lower semicontinuity 
occurs  also
for 
$(Y^{1-}_t)_{t\geq1}$
in the case where
$-1\in\cD_\Lambda^\circ$.
\smallskip

\noindent \textbf{(b)} Lemma~\ref{lem:perturbation} implies that if
$(Z_t)_{t\geq1}$
satisfies the assumptions of Theorem~\ref{thm:GartnerEllis}
\textit{and}
$\cD_\Lambda$
is contained in
$(-\infty,\lambda)$,
for some
$\lambda>0$,
then 
$(Y_t^{\lambda+})_{t\geq1}$
also satisfies the assumptions of Theorem~\ref{thm:GartnerEllis}
and hence the LDP with a good rate function
$\Lambda^*$.
An analogous statement holds for
$(Y_t^{\lambda-})_{t\geq1}$.
\end{remarks}

\begin{proof}
Note that 
$\log \EE\left[\E^{uE_\lambda}\right]$
is 
finite and equal to 
$\log\left(\lambda/(\lambda-u)\right)$
if and only if
$u\in(-\infty,\lambda)$.
For all large 
$t$
and
$u\in\mathcal D_\Lambda\cap(-\infty,\lambda)$,
the assumption in~\eqref{eq:LDP_Assumption}
implies that 
$\Lambda^{\lambda+}_t(tu)=\log \EE\left[\exp\left(tuY_t^{\lambda+}\right)\right]$
is finite 
and that the formula holds
\begin{eqnarray}
\label{eq:Perturbed_Cum_Gen_Fun}
\Lambda^{\lambda+}_t(tu) = \Lambda_t(tu)+\log\frac{\lambda}{\lambda-u},
\qquad\text{where}\qquad
\Lambda_t(tu) = \log \EE\left[\exp\left(tuZ_t\right)\right].
\end{eqnarray}
%\qquad\text{if}\qquad
%u\in\mathcal D_\Lambda\cap(-\infty,\lambda).
The inequality 
$u\geq\lambda$
implies that,
since 
$\Lambda_t(tu)>-\infty$,
we have 
$\Lambda^{\lambda+}_t(tu)=\infty$
for all 
$t$
and hence
$\Lambda^{\lambda+}(u)=\infty$.
If 
$u\in(\bbR\setminus\cD_\Lambda)\cap(-\infty,\lambda)$,
then~\eqref{eq:Perturbed_Cum_Gen_Fun}
yields
$\Lambda^{\lambda+}(u)=\lim_{t\nearrow\infty}\Lambda^{\lambda+}_t(tu)/t =
\infty$.
This proves the lemma for 
$(Y^{\lambda+}_t)_{t\geq1}$.
The case of 
$(Y^{\lambda-}_t)_{t\geq1}$
is analogous.
\end{proof}

%========================================================================================
\section{Essential smoothness can fail}
\label{sec:Trap}

The Heston model 
$S=\E^{X}$
is a stochastic volatility model
with the log-stock
process 
$X$
given by
%\begin{eqnarray}
%\dd X_t & = & -\frac{1}{2} Y_t\dd t+\sqrt{Y_t}\dd W^1_t\\
%\dd Y_t & = & \kappa(\theta-Y_t)\dd t+\sigma \sqrt{Y_t}\dd W^2_t
%\end{eqnarray}
\begin{eqnarray}
\label{eq:Heston_SDE}
\dd X_t  =  -\frac{Y_t}{2} \dd t+\sqrt{Y_t}\dd W^1_t &\text{and} & 
\dd Y_t  =  \kappa(\theta-Y_t)\dd t+\sigma \sqrt{Y_t}\dd W^2_t,
\end{eqnarray}
where
$\kappa,\theta,\sigma>0$,
$Y_0=y_0>0$,
$X_0=x_0\in\bbR$
and
$W^1,W^2$
are standard Brownian motions 
with correlation 
$\rho\in(-1,1)$.
%For simplicity and without loss of generality 
%let the price of the risky security
%$S$
%at time zero be normalized to one 
%(i.e. 
%$X_0=0$).
The standing assumption
\begin{eqnarray}
\label{eq:assumption_chi}
%\chi(1)<0, &\text{where} & 
%\chi(u)=u
\rho\sigma-\kappa<0,
\end{eqnarray}
is made in~\cite{FJ09Large} 
(see equation~(2.2) in Theorem~2.1
on page 5 of~\cite{FJ09Large}).
In particular
the inequality in~\eqref{eq:assumption_chi}
implies that 
$S$
is a strictly positive true martingale 
%(see e.g.~\cite[Corollary~2.7(b)]{K2008a})
and allows the definition of the share measure
$\wt \PP$
%defined after formula~\eqref{eq:Call_Formulae}
via the Radon-Nikodym derivative 
$\dd \wt \PP/\dd \PP = \E^{X_t-x_0}$.

The authors'
aim in~\cite{FJ09Large} is to obtain the
limiting implied volatility smile 
as maturity tends to infinity 
at the %(maturity dependent) 
strike
$K=S_0\E^{xt}$
for any
$x\in\bbR$
in the Heston model. %under the assumptions above, 
Their main formula is given in
Corollary~3.1~of~\cite{FJ09Large}.
A key step in the proof of~\cite[Corollary~3.1]{FJ09Large}
is given by~\cite[Corollary~2.4]{FJ09Large}.
%In~\cite{FJ09Large}
%this problem  is tackled in the following three steps:
%\begin{enumerate}[(I)]
%\item 
%\label{enum:Step_1}
%establish the LDP for 
%$(X_t/t)_{t\geq1}$
%under 
%$\PP$ (see \cite[Theorem~2.1]{FJ09Large})
%and under the share measure 
%$\wt \PP$ (see~\cite[Corollary~3.1]{FJ09Large}); 
%\item 
%\label{enum:Step_2}
%apply formulae~\eqref{eq:Put_Formulae} and~\eqref{eq:Call_Formulae} to the Heston model
%at the strike
%$K=S_0\E^{xt}$
%and use~\eqref{enum:Step_1}
%to obtain the limits of the vanilla option prices
%(see~\cite[Corollary~2.4]{FJ09Large});
%\item 
%\label{enum:Step_3}
%apply~\eqref{enum:Step_2} to find the limit of the implied
%volatility smile (see~\cite[Corollary~2.14]{FJ09Large}).
%\end{enumerate}
%A simple proof of
%step~\eqref{enum:Step_1}
%is given in Remarks~(i) and~(ii)
%below. 
%Step~\eqref{enum:Step_3} can be carried out
%as described in the 
%proof of~\cite[Corollary~2.14]{FJ09Large} 
%once the limiting formulae for the vanilla option
%prices of~\eqref{enum:Step_2} have been established. 
%It is the link between~\eqref{enum:Step_1}
%and~\eqref{enum:Step_2} and the proof of step~\eqref{enum:Step_2}
%that constitute the main focus of this note.
%Furthermore the validity of~\eqref{enum:Step_2} 
%is crucial in establishing 
%the main result~\cite{FJ09Large}
%(i.e. Corollary~3.1~in \cite{FJ09Large}).
In the proof of~\cite[Corollary~2.4]{FJ09Large}
(see last line on page 17 and lines 4 and 14 on page 18)
it is implicitly assumed that 
%asserted that
%formulae~\eqref{eq:Put_Formulae} and~\eqref{eq:Call_Formulae} 
%can be used to find the limits of the vanilla option prices 
%by applying the LDP to the families of random variables
%$(X_t/t\pm E_1/t)_{t\geq1}$
%to obtain the limits of the tail probabilities.
%In other words the authors of~\cite{FJ09Large}
%apply the ``fact''
%that~\eqref{enum:Step_1} implies 
the LDP for
$(X_t/t_{t\geq1}$
implies the LDP for the family
$(X_t/t\pm E_1/t)_{t\geq1}$.
%and hence, by formulae~\eqref{eq:Put_Formulae} and~\eqref{eq:Call_Formulae}, 
%establishes the limits in~\eqref{enum:Step_2}.
However, as we have seen in Section~\ref{sec:trap} (see remarks
following Lemma~\ref{lem:perturbation}), 
Theorem~\ref{thm:GartnerEllis} cannot be applied directly
to the family 
$(X_t/t\pm E_1/t)_{t\geq1}$,
even if 
$(X_t/t)_{t\geq1}$
satisfies its assumptions.
%In this note we show how to circumvent this problem
%(see~Theorem~\ref{thm:Fix}  below).
%avoids this problem
%Hence further work %, presented in 
%is required to establish the limits in~\eqref{enum:Step_2}.
We start with a precise description of the problem
and present the solution in
Theorem~\ref{thm:Fix}. 

\begin{remarks}
\noindent \textbf{(i)}
Under~\eqref{eq:assumption_chi},
%In~\cite{FJ09Large} the authors assume that
%$\chi(1)<0$
%(see equation~(2.2) in Theorem~2.1
%on page 5 of~\cite{FJ09Large}),
%which implied that 
%$S$
%is a true martingale (see e.g.~\cite[Corollary~2.7(b)]{K2008a}) 
%and yields by
a simple calculation implies 
that
%a direct application of~\cite[Theorem~3.4]{K2008a} yields an explicit 
%formula for the limiting cumulant generating function 
$\Lambda$
and %the effective domain 
$\cD_\Lambda$
of the family %the family %of random variables
$(X_t/t)_{t\geq1}$
are:
%take the form:
\begin{eqnarray}
\label{eq:Lmabda}
\Lambda(u) & = & -\frac{\theta\kappa}{\sigma^2}\left(u\rho\sigma-\kappa+\sqrt{\Delta(u)}\right)\quad \text{for}\quad
u\in\cD_\Lambda \qquad\text{and}\qquad \cD_\Lambda = [u_-,u_+]\quad\text{where}\\
u_\pm & = & \left(1/2-\rho\kappa/\sigma
\pm\sqrt{\left(\kappa/\sigma-\rho\right) \kappa/\sigma+1/4}\right)/\left(1-\rho^2\right)
\quad\text{with}\quad u_-<0<1<u_+.
%\pm\left[(1/2-\kappa/\sigma)^2+(1-\rho)\kappa/\sigma\right]^{1/2}\right)/\left(1-\rho^2\right).
\label{eq:roots}
\end{eqnarray}
In~\eqref{eq:Lmabda}
the function
$\Delta$
is a quadratic 
$\Delta(u)=(u\rho\sigma-\kappa)^2-\sigma^2(u^2-u)$
and the boundary points
$u_+$ and $u_-$
of the effective domain
$\cD_\Lambda$
are its zeros.
Elementary calculations show that 
$\Lambda$
is essentially smooth 
and that the unique minimum of
$\Lambda^*$
is %by~\eqref{eq:Lmabda}
attained at
$\Lambda'(0)=-\theta/2$.
%It is easy to see that~\eqref{eq:roots}
%and
%assumptions 
%$\chi(1)<0$
%and
%$\kappa>0$
%yield
%$u_+>1$
%and
%$u_-<0.$
%Furthermore the formula 
%in~\eqref{eq:Lmabda}
%implies 
%the essential smoothness 
%and lower semicontinuity 
%of
%$\Lambda$.
Therefore 
$(X_t/t)_{t\geq1}$
satisfies the LDP with the good rate function
$\Lambda^*$,
defined in~\eqref{eq:DefFenchelLegendreTransf},
by Theorem~\ref{thm:GartnerEllis}.
%Since 
%$\Lambda$
%is differentiable on
%$\cD_\Lambda^\circ$
%and
%$\Lambda'$
%is strictly increasing, 
%the definition in~\eqref{eq:DefFenchelLegendreTransf}
%implies that 
%the transform 
%$\Lambda^*$
%is also differentiable with strictly increasing 
%derivative
%$(\Lambda^*)'=(\Lambda')^{-1}$.
%The unique minimum of
%$\Lambda^*$
%is by~\eqref{eq:Lmabda}
%attained at
%$\Lambda'(0)=-\theta/2$.
%This gives an alternative short proof of~\cite[Theorem~2.1]{FJ09Large}.
\smallskip

\noindent \textbf{(ii)} 
Under the share measure
$\wt \PP$,
given by
%(with Radon-Nikodym derivative 
$\dd \wt \PP/\dd \PP = \E^{X_t-x_0}$,
we have 
$\wt \EE\left[\E^{u X_t}\right]= \E^{-x_0}\EE\left[\E^{(u+1) X_t}\right]$
for all 
$u\in\bbR$
and
$t>0$
and hence the family %of random variables
$(X_t/t)_{t\geq1}$
under 
$\wt \PP$
satisfies the assumption in~\eqref{eq:LDP_Assumption}
with the limiting cumulant generating function 
%$\wt \Lambda$
%of the form
%in~\eqref{eq:LDP_Assumption}
%with the 
$\wt \Lambda(u)=\Lambda(u+1)$,
$\cD_{\wt \Lambda}=[u_--1,u_+-1]$.
%Definition~\eqref{eq:DefFenchelLegendreTransf}
%yields
As before,
$(X_t/t)_{t\geq1}$
satisfies the LDP 
under 
$\wt \PP$
with the strictly convex good rate function
$\wt \Lambda$,
which satisfies
%Furthermore
%$\wt \Lambda$
%is essentially smooth and lower semicontinuous, 
%satisfies
$\wt \Lambda^*(x)=\Lambda^*(x)-x$
for all 
$x\in\bbR$
and
attains its unique minimum at 
$\wt \Lambda'(0)=\Lambda'(1)=\theta\kappa/(\kappa-\rho\sigma)$.
%Furthermore~\eqref{eq:Lmabda}
%and~\eqref{eq:roots}
%imply that 
%$\wt \Lambda$
%is essentially smooth and lower semicontinuous. 
%Hence,
%Theorem~\ref{thm:GartnerEllis}
%implies that
%$\wt \Lambda^*$.
%that 
%attains its unique minimum at 
%$\wt \Lambda'(0)=\Lambda'(1)=\theta\kappa/(\kappa-\rho\sigma)$.
%This remark contains the statement and proof 
%of~\cite[Corollary~3.1]{FJ09Large}.
%\smallskip
%
%\noindent \textbf{(iii)} The authors of~\cite{FJ09Large}
%cite the work by Keller-Ressel~\cite{K2008a} but do not apply
%Theorem~3.4 in~\cite{K2008a} to prove~\cite[Theorem~2.1]{FJ09Large}
%as was done in Remark~(i) above. Their argument (see pages 12 to 15
%in~\cite{FJ09Large} for the proof of Theorem~2.1) is based 
%on the explicit closed form 
%of the cumulant generating function
%$\Lambda_t$
%at a finite time
%$t$, 
%which is known in the case of the Heston model.
\end{remarks}

%We now state our main result.

\begin{theorem}
\label{thm:Fix}
%Let 
%$S=\E^X$
%be the Heston model with 
Let the process 
$X$
be given by~\eqref{eq:Heston_SDE}
and assume that~\eqref{eq:assumption_chi}
holds.
%is satisfied. 
Let 
$E_1$
be the exponential random variable with
$\EE[E_1]=1$,
which is independent of 
$X$.
Then the following limits hold:
\begin{eqnarray}
\label{eq:lim_1}
\lim_{t\nearrow\infty}\frac{1}{t} \log \PP\left[X_t-x_0+E_1<xt\right] & = &
-\Lambda^*(x)
\qquad\text{for}\quad x\leq \Lambda'(0)=-\theta/2;\\
\label{eq:lim_2}
\lim_{t\nearrow\infty}\frac{1}{t} \log \wt \PP\left[X_t-x_0-E_1>xt\right] & = & x-\Lambda^*(x)
\qquad\text{for}\quad x\geq \Lambda'(1)=\theta\kappa/(\kappa-\rho\sigma);\\
\label{eq:lim_3}
\lim_{t\nearrow\infty}\frac{1}{t} \log \wt \PP\left[X_t-x_0-E_1\leq xt\right] & = & x-\Lambda^*(x)
%\qquad\text{for}\quad x\in\left[\theta/2,\theta\kappa/(\kappa-\rho\sigma)\right];
\qquad\text{for}\quad x\in\left[\Lambda'(0),\Lambda'(1)\right];
\end{eqnarray}
where 
$\Lambda$
is given in~\eqref{eq:Lmabda},
its Fenchel-Legendre transform 
$\Lambda^*$
is defined in~\eqref{eq:DefFenchelLegendreTransf}
and
$\dd \wt\PP/\dd \PP=\E^{X_t-x_0}$.
\end{theorem}

\begin{remark}
\noindent  The limits in Theorem~\ref{thm:Fix}
are precisely the limits that arise in the proof 
of~\cite[Corollary~2.4]{FJ09Large}
(see the last line on page~17 and lines~4 and~14 on page~18)
and are claimed to hold since the family
$(X_t/t)_{t\geq1}$
satisfies the LDP 
under 
$\PP$
and
$\wt\PP$
by Remarks~(i) and~(ii) above and Theorem~\ref{thm:GartnerEllis}. 
However 
Lemma~\ref{lem:perturbation} 
implies that the 
limiting cumulant generating function 
$\Lambda^{1+}$
of the family of random variables 
$(Z_t+E_1/t)_{t\geq1}$,
where
$Z_t=(X_t-x_0)/t$,
is neither lower semicontinuous nor essentially smooth.
%the effective domain 
%of 
%$(Z_t)_{t\geq1}$
%as 
%$\cD_\Lambda=[u_-,u_+]$
%and Remark~(i)
%%~\eqref{eq:roots} 
%yields that 
%$u_+>1$.
%Lemma~\ref{lem:perturbation} 
%implies that the 
%limiting cumulant generating function 
%$\Lambda^{1+}$
%of 
%$(Z_t+E_1/t)_{t\geq1}$
%is neither lower semicontinuous nor essentially smooth
%and 
Hence
Theorem~\ref{thm:GartnerEllis} 
cannot be applied  to
%to the family of random variables
$(Z_t+E_1/t)_{t\geq1}$.
An anologous issue arises under the measure
$\wt\PP$.
%\noindent \textbf{(C)} In~\eqref{eq:lim_2} and~\eqref{eq:lim_3} 
%the limiting behaviour of the tail probabilities of the 
%family of random variables 
%$(Z_t-E_1/t)_{t\geq1}$
%under
%$\wt\PP$,
%where
%$Z_t=(X_t-x_0)/t$,
%is of interest.
%Remark~(ii) above implies that the effective domain of
%$(Z_t)_{t\geq1}$
%under
%$\wt\PP$
%is 
%$\cD_{\wt \Lambda}=[u_--1,u_+-1]$.
%Since Remark~(i)
%%~\eqref{eq:roots} 
%yields
%$u_-<0$,
%Lemma~\ref{lem:perturbation}
%implies that the limiting cumulant generating function of
%$(Z_t-E_1/t)_{t\geq1}$
%under
%$\wt\PP$
%is neither lower semicontinuous nor essentially smooth
%for any parameter set.
%Therefore the assumptions of 
%Theorem~\ref{thm:GartnerEllis} 
%are not satisfied.
\end{remark}

\begin{proof}
The basic idea of the proof is simple: for~\eqref{eq:lim_1}
we sandwich the probability 
$\PP\left[X_t-x_0+E_1<xt\right]$
between two tail probabilities of two families of random variables,
which satisfy the LDP 
with the same rate function
$\Lambda^*$ 
by Lemma~\ref{lem:perturbation} and Theorem~\ref{thm:GartnerEllis}.
%to conclude that these two families satisfy the LDP 
%with the rate function
%$\Lambda^*$. 
The limits in
\eqref{eq:lim_2} and~\eqref{eq:lim_3}
follow similarly. 

For given parameter values in the Heston model
%$S=\E^X$
pick
$\lambda>u_+$,
where
$u_+$
is defined in~\eqref{eq:roots}.
Let 
$E_\lambda$
be an exponential random variable
with
$\EE[E_\lambda]=1/\lambda$,
defined on the same probability space as
$X$
and
$E_1$
and independent of both.
Since 
$u_+>1$,
we have the elementary inequality
\begin{eqnarray}
\label{eq:ElemIneq}
\PP\left[E_\lambda<\alpha\right]=
I_{\{\alpha>0\}}\left(1-\E^{-\lambda \alpha}\right)\leq
I_{\{\alpha>0\}}\left(1-\E^{-\alpha}\right)=
\PP\left[E_1<\alpha\right]\qquad\text{for
any}\quad\alpha\in\bbR.
\end{eqnarray}
The inequality
\begin{eqnarray}
\label{eq:MainInequality}
\PP\left[X_t-x_0+E_\lambda<xt\right] & \leq & 
\PP\left[X_t-x_0+E_1<xt\right]
\end{eqnarray}
follows 
by conditioning on 
$X_t$
and 
applying~\eqref{eq:ElemIneq}. 
On the other hand, since
$E_1>0$
a.s., we have
\begin{eqnarray}
\label{eq:SimpleMainInequality}
\PP\left[X_t-x_0+E_1<xt\right]  & \leq & 
\PP\left[X_t-x_0<xt\right]. 
\end{eqnarray}
Lemma~\ref{lem:perturbation} 
implies that the families of random variables
$(Z_t+E_\lambda/t)_{t\geq1}$
and
$(Z_t)_{t\geq1}$,
where
$Z_t=(X_t-x_0)/t$,
both have the limiting cumulant generating function equal to
$\Lambda$
given in~\eqref{eq:Lmabda}
with the effective domain 
$\cD_\Lambda=[u_-,u_+]$.
Since 
$\Lambda$
is essentially smooth and lower semicontinuous on
$\cD_\Lambda$
and the assumption in~\eqref{eq:LDP_Assumption}
is satisfied, Theorem~\ref{thm:GartnerEllis}
implies that 
$(Z_t+E_\lambda/t)_{t\geq1}$
and
$(Z_t)_{t\geq1}$,
satisfy the LDP with the good rate function
$\Lambda^*$.
Since 
$x$
in~\eqref{eq:lim_1}
is assumed to be less or equal to the unique minimum
$\Lambda'(0)=-\theta/2$
of 
$\Lambda^*$
(see Remark~(i) above)
and
$\Lambda^*$
is non-negative and strictly convex, the LDP (see the inequalities in~\eqref{eq:DefLDP})
and the inequalities in~\eqref{eq:MainInequality} and~\eqref{eq:SimpleMainInequality}
imply the limit in~\eqref{eq:lim_1}.

To prove~\eqref{eq:lim_2}
pick
$\lambda>1-u_-$
and note that
%Remark~(i)
%implies the inequality
%$\lambda>1$.
%Therefore 
the inequality in~\eqref{eq:ElemIneq}
and conditioning on 
$X_t$
yield
\begin{eqnarray}
\label{eq:TildeInequality}
\wt \PP\left[X_t-x_0>xt\right] \geq
\wt \PP\left[X_t-x_0-E_1>xt\right] \geq
\wt \PP\left[X_t-x_0-E_\lambda>xt\right]. 
\end{eqnarray}
As before,
Lemma~\ref{lem:perturbation} 
and Theorem~\ref{thm:GartnerEllis}
imply 
that 
$(Z_t-E_\lambda/t)_{t\geq1}$
and
$(Z_t)_{t\geq1}$
satisfy the LDP 
with the convex rate function
$\wt \Lambda^*$,
which 
by Remark~(ii) above
attains its unique minimum
at
$\Lambda'(1)=\theta\kappa/(\kappa-\rho\sigma)$.
Since 
$x\geq\Lambda'(1)$
in~\eqref{eq:lim_2},
the limit follows.
%under
%$\wt\PP$
%have the limiting cumulant generating function 
%$\wt \Lambda(u)=\Lambda(u+1)$,
%where
%$\Lambda$ 
%is given in~\eqref{eq:Lmabda},
%with the effective domain 
%$\cD_{\wt\Lambda}=[u_--1,u_+-1]$
%(see Remark~(ii) above).
%Furthermore
%$\wt \Lambda$
%is essentially smooth and lower semicontinuous.
%Hence under
%$\wt\PP$
%the families
%$(Z_t-E_\lambda/t)_{t\geq1}$
%and
%$(Z_t)_{t\geq1}$
%satisfy 
%the LDP 
%with the rate function
%$\wt \Lambda^*$
%by Theorem~\ref{thm:GartnerEllis}.
%By Remark~(ii) above,
%$\wt \Lambda^*$
%is strictly convex, attains its unique minimum
%at 
%$\Lambda'(1)=\theta\kappa/(\kappa-\rho\sigma)$
%and satisfies 
%$\wt \Lambda^*(x)=\Lambda^*(x)-x$.
%Since 
%$x\geq\Lambda'(1)$
%in~\eqref{eq:lim_2},
%the limit follows.
%Note that 
A similar argument implies the limit in~\eqref{eq:lim_3}
for all 
$x\in[\Lambda'(0),\Lambda'(1)]$, 
which concludes the proof.
%$\Lambda'$
%is strictly increasing and hence
%$\Lambda'(0)<\Lambda'(1)$. 
%The probabilities under
%$\wt \PP$
%of the complements of the events in~\eqref{eq:TildeInequality}
%yield inequalities which imply the limit in~\eqref{eq:lim_3}
%for all 
%$x\in[\Lambda'(0),\Lambda'(1)]$. 
%This concludes the proof.
\end{proof}

\begin{comment}
%========================================================================================
\section{Conclusion}
\label{sec:Conclusion}

This note investigates the subtleties that arise with effective domains
and a potential loss of essential smoothness when a family of random variables
is perturbed by an independent exponential (the precise formulation is given in
Section~\ref{sec:trap}). 
A gap in the proof of Corollary~2.4 in~\cite{FJ09Large}, which is key in deriving
the main result of~\cite{FJ09Large}, is identified and circumvented by 
Theorem~\ref{thm:Fix}. The proof of Theorem~\ref{thm:Fix} 
presented in Section~\ref{sec:Trap}
has two main features: one,
it is based on the relationship between the essential smoothness 
of large deviation theory and option pricing formulae~\eqref{eq:Put_Formulae} 
and~\eqref{eq:Call_Formulae} developed in this note and two, it
stays well within the realm of~\cite{FJ09Large} without the need for the introduction
of further mathematical concepts.
\end{comment}

%========================================================================================

%\bibliographystyle{alpha}
%\bibliography{references}

\begin{thebibliography}{9}

%\bibitem{CM09}P.~Carr and D.~Madan.
%Saddlepoint Methods for Option Pricing.
%\textit{Journal of Computational Finance} {\tt 13} (1): 49-61, 2009.

\bibitem{DemboZeitouni}A.~Dembo and O.~Zeitouni.
\textit{Large Deviations Techniques and Applications}.
Springer-Verlag, New York, 2nd ed., 1998.
%\textit{Jones and Bartlet publishers, Boston}, 1998.

\bibitem{FJ09Large}M.~Forde and A.~Jacquier.
The large-maturity smile for the Heston model.
To appear in \textit{Finance \& Stochastics}. 


%\bibitem{K2008a}M.~Keller-Ressel,
%Moment explosions and long-term behavior of affine stochastic volatility
%models.  arXiv:0802.1823, 2008. Forthcoming in \textit{Mathematical Finance}.


%\bibitem{Rockafellar}A.~T.~Rockafellar.
%\textit{Convex Analysis}.
%Princeton Landmarks in Mathematics, Princeton University Press,
%New Jersey, 1970.

\end{thebibliography}

\end{document}